\documentclass[prd,aps]{revtex4}
\usepackage{amsmath}
\usepackage{amssymb}
\usepackage{amsfonts}
\usepackage{graphicx,bm}
\usepackage{dcolumn}
\usepackage[colorlinks=true]{hyperref}
\usepackage{epsf}
\usepackage{enumerate}
\usepackage{hhline}
\usepackage{array}
\usepackage{tabularx}
\usepackage{subfigure}

\newcommand{\be}{\begin{equation}}
\newcommand{\ee}{\end{equation}}
\newcommand{\bea}{\begin{eqnarray}}
\newcommand{\eea}{\end{eqnarray}}
\newcommand{\beaa}{\begin{eqnarray*}}
\newcommand{\eeaa}{\end{eqnarray*}}

\newcommand{\e}{\mathrm{e}}

\def\be{\begin{equation}}
\def\ee{\end{equation}}
\def\bea{\begin{eqnarray}}
\def\eea{\end{eqnarray}}

\begin{document}
\title{Extremal cosmological black holes in Horndeski gravity and the anti-evaporation regime}
\author{Ismael Ayuso}
\email{iayuso@fc.ul.pt} \affiliation{Departamento de F\'isica and Instituto de Astrof\'isica e Ci\^encias do Espa\c{c}o, Faculdade de Ci\^encias,\\ Universidade de Lisboa, Edif\'icio C8, Campo Grande, 1769-016 Lisboa, Portugal} 
\author{Diego S\'aez-Chill\'on G\'omez}
\email{diego.saez@uva.es} \affiliation{Department of Theoretical Physics, Atomic and Optics, Campus Miguel Delibes, \\ University of Valladolid UVA, Paseo Bel\'en, 7,
47011 - Valladolid, Spain}

\begin{abstract}
Extremal cosmological black holes are analysed in the framework of the most general second order scalar-tensor theory, the~so-called Horndeski gravity. Such~extremal black holes are a particular case of Schwarzschild-De Sitter black holes that arises when the black hole horizon and the cosmological one coincide. Such~metric is induced by a particular value of the effective cosmological constant and is known as Nariai spacetime. The~existence of this type of solutions is studied when considering the Horndeski Lagrangian and its stability is analysed, where the so-called anti-evaporation regime is studied. Contrary to other frameworks, the~radius of the horizon remains stable for some cases of the Horndeski Lagrangian when considering perturbations at linear order.

\end{abstract}
%
%
\maketitle
%
%
%
\section{Introduction}
General Relativity (GR) has shown its power of prediction over more than one hundred years, and~despite some important issues, it~is still considered as the best description of gravity. Nevertheless, there are some fundamental questions to be answered in the future in the context of theoretical physics. From~a UV completion of gravity to cosmological late-time acceleration, among other also relevant problems, the~scientific community is making a great effort to afford them. In~particular, black hole solutions have been widely studied in the literature, as~are natural solutions of GR and have led to an important development of gravitational physics, including the famous theorems about singularities that offer a way to understand  these objects and their main features better. Particularly Schwarzschild-(Anti) De Sitter spacetime arises in GR as a solution when considering a (negative) cosmological constant. The~black holes described by this spacetime have been of great interest as they show a thermodynamical equilibrium when analysing Hawking radiation~\cite{Gibbons-Hawking,Gibbons-Hawking2,Hawking:1982dh}. In~the case of a positive cosmological constant, the~Schwarzschild-de Sitter spacetime shows in general two horizons, one~corresponding to the black hole event horizon and the other one to a cosmological horizon. The~extreme case arises when both horizons coincide at the same hypersurface, the~so-called Nariai spacetime~\cite{Nariai}, leading to an interesting structure for the spacetime and the trajectories of geodesics~\cite{Podolsky:1999ts} as well as for its spectrum~\cite{MaassenvandenBrink:2003yq}. In~addition, the~stability of such extreme spacetime has been studied in~\cite{Ginsparg}, but~when some corrections are included, an~interesting phenomenon occurs, as~the radius of the horizon becomes unstable and grows, what~has been called black hole anti-evaporation~\cite{Bousso:1997wi}. Despite that the anti-evaporation regime was initially studied and attributed to semiclassical corrections that affect the evaporation of black holes in de Sitter spacetime when analysing the one-loop effective action~\cite{Bousso:1997wi,Nojiri:1998ue}, other frameworks that lead to classical instabilities that affect the radius of the horizon have been also named  antievaporation, as~the case of $F(R)$ gravity~\cite{Nojiri:2013su}, Gauss--Bonnet gravities~\cite{Sebastiani:2013fsa}, bigravity theories~\cite{Katsuragawa:2015sjn,Katsuragawa:2014hda} and mimetic gravity~\cite{Nashed:2018aai}.\par

In the context of cosmology, the~main issue lies on the unknown dark energy (also on dark matter), which has been widely contrasted by observational data and many theoretical models have been proposed to explain its main consequence, the~late-time acceleration of the universe expansion (for a review see~\cite{reviews,Huterer:2017buf,Frieman:2008sn,Heisenberg:2018vsk,Joyce:2014kja}). Some~of such dark energy models are focused on modifications of GR, which may provide a natural solution to the problem, which might be connected to the corrections expected from some UV completions of GR, such as~string theory~\cite{Nojiri:2006je}. In~this sense, the~simplest way of modifying GR is by introducing a scalar field, which  incorporates an additional scalar mode while keeping the well known predictions by GR unbroken through screening mechanism that can be implemented by an appropriate potential---the chameleon mechanism---and by the kinetic term---the Vainshtein mechanism. In~addition, scalar-tensor theories are well known and well understood, from~the Brans--Dicke theory to Horndeski gravity, there is a wide range of scalar field models that have been widely analysed and used not only to provide a natural explanation for dark energy but also to get a better understanding of GR itself~\cite{Elizalde:2008yf}. Generalisations of standard scalar-tensor theories have been widely studied lately, mainly in the context of cosmology, as~the so-called K-essence, which presents a non-canonical kinetic term and provides a natural explanation for dark energy~\cite{ArmendarizPicon:2000dh,ArmendarizPicon:2000ah}, or~the so-called Galileons, that~incorporates a Galilean-like symmetry and which can also reproduce in a simple way the late-time acceleration~\cite{Nicolis:2008in}. These~types of models have in common that they may avoid the so-called Ostrogradsky instability that arises in higher order theories, which is absent in second order theories, such as the ones cited above. This~class of scalar-tensor models are encompassed in the so-called Horndeski gravity~\cite{Horndeski:1974wa}, which represents the most general theory with second order field equations (for a review see~\cite{Kobayashi:2019hrl,Deffayet:2013lga}). Horndeski gravity is shown to be a generalisation of Galileon in its covariant form~\cite{Deffayet:2009wt}, which is also connected to k-essence fields~\cite{Deffayet:2011gz}. Nevertheless, there have been some healthy extensions of Horndeski gravity also implying second order derivatives for the field equations~\cite{Zumalacarregui:2013pma,Achour:2016rkg,Gleyzes:2014dya}. In~general, Horndeski gravity is well understood in many contexts, inflationary models have been widely analysed as well as the growth of cosmological perturbations~\cite{Kobayashi:2011nu,DeFelice:2011uc,Gao:2011qe,DeFelice:2011hq,Gao:2011vs}, also~consequently dark energy models can be easily implemented in Horndeski gravity~\cite{Charmousis:2011bf,Copeland:2012qf}, the predictions and restrictions of which are analysed~\cite{Kobayashi:2016xpl,Amendola:2012ky,Zumalacarregui:2016pph,Ezquiaga:2017ekz}. Also~in light of the era of gravitational waves~\cite{TheLIGOScientific:2017qsa,Monitor:2017mdv}, Horndeski gravity is shown to carry just an additional scalar
 mode~\cite{Deffayet:2015qwa}, but~the theory is well constrained by the speed of propagation of the graviton ~\cite{Bettoni:2016mij,Deffayet:2010qz,Bartolo:2017ibw}, which implies several restrictions on the full Lagrangian~\cite{Kase:2018aps}.

Also, static spherically symmetric solutions, such as~black holes, have~been widely studied in the literature within theories beyond GR~\cite{delaCruzDombriz:2009et,Olmo:2011ja,Olmo:2013gqa,delaCruzDombriz:2012xy,Sotiriou:2013qea,Clifton:2006ug,Yunes:2007ss,Cardoso:2009pk}, as~they may provide a way to regularise such types of solutions~\cite{Olmo:2015bya,Olmo:2015dba,Bejarano:2017fgz,Nojiri:2017kex}, a~better understanding of Birkhoff's theorem~\cite{Faraoni:2010rt,Schleich:2009uj,Capozziello:2011wg}, or~ new direct ways for testing General Relativity~\cite{Moffat:2015kva,Davis:2014tea,Guo:2018kis}. In~Horndeski gravity, there have been plenty of works where such types of solutions are studied, mainly when dealing with compact objects such as black holes~\cite{Babichev:2017guv,Tattersall:2018nve,Babichev:2016rlq,Babichev:2016fbg}, but~also when assuming the constraints imposed on the full Horndeski Lagrangian by the speed of propagation of gravitational waves~\cite{Tattersall:2018map}, and~the stability of such types of spacetimes ~\cite{Babichev:2017lmw,Ganguly:2017ort,Sakstein:2016oel,Ogawa:2015pea,Babichev:2016rlq,Babichev:2016fbg}. The~no-hair theorem is also extended in these theories~\cite{Lehebel:2017fag}. Moreover, the~Cauchy problem has been analysed in Horndeski gravity by studying the hyperbolicity of the system of equations, which seems to admit a well posed initial value problem~\cite{Kovacs:2020ywu}. Also~the stability in non perturbative cosmology has been studied in~\cite{Ijjas:2018cdm} as well as the gauge problem in such Lagrangians~\cite{Ijjas:2017pei}. 

The aim of the present paper is to analyse Nariai spacetime in Horndeski gravity and the emergence of the anti-evaporation regime by studying the corresponding perturbations on the metric. Perturbations in Schwarzschild black holes and the Cauchy problem have been widely analysed in the literature within several gravitational theories~\cite{Martel:2005ir,Kimura:2018nxk,Hung:2017qop,Cardoso:2009pk}. Here~we intend to describe how perturbations of a scalar field around a constant background value can affect the radius of the horizon of the black hole. To~do so, we~study the existence of Schwarzschild-de Sitter solutions in its extremal version for the Lagrangians that compose Horndeski gravity, which also show some implications on an extended version of Birkhoff's theorem for scalar-tensor theories. Finally, we~analyse the stability of such solution for a shorter version of the full Horndeski Lagrangian, motivated by keeping as few free functions as possible and which coincides with the viable terms restricted by the speed of GW's. 

The paper is organised as follows: In Section \ref{background}, a brief introduction to Horndeski gravity and the Nariai metric is provided. Section \ref{solutions} is devoted to the analysis of the viable Lagrangians that contain the Nariai metric as a solution. In~Section \ref{antievaporation}, the~anti-evaporation regime is analysed. Finally, Section \ref{conclusions} gathers the conclusions.

\section{Nariai Spacetime in Horndeski Gravity}
\label{background}

Let us start by writing the general action that we are dealing with throughout this manuscript. This~is the Hilbert--Einstein action plus the so-called Horndeski Lagrangian:
\be
S_G=\int dx^4 \sqrt{-g} \left[\frac{R}{16\pi G}+\mathcal{L}_{\text{Hr}}+\mathcal{L}_m\right]\ ,
\label{Action}
\ee  
where $\mathcal{L}_m$ is the matter Lagrangian, which encompasses all the matter species of the system under study while the Horndeski Lagrangian $\mathcal{L}_{\text{Hr}}$ is given by:
\begingroup\makeatletter\def\f@size{9}\check@mathfonts
\def\maketag@@@#1{\hbox{\m@th\normalsize \normalfont#1}}%
\be
\begin{array}{rrl}
\mathcal{L}_{\text{Hr}}&=& G_2(\phi, X)-G_3(\phi, X)\Box\phi+G_4(\phi, X)R+G_{4X}(\phi, X) \left[(\Box\phi)^2-\phi_{;\mu\nu}\phi^{;\mu\nu}\right]+G_5(\phi, X)\phi_{;\mu\nu}G^{\mu\nu} \\
&&-\frac{G_{5X}(\phi, X)}{6}\left[(\Box\phi)^3-3\Box\phi\phi_{;\mu\nu}\phi^{;\mu\nu}+2\phi_{;\mu\nu}\phi^{;\nu\lambda}\phi^{;\mu}_{\lambda}\right]\ .
\label{Horndeski}
\end{array}
\ee
\endgroup

Here, $\phi$ is a scalar field, $G_{\mu\nu}$ is the Einstein tensor, $_{;\mu}=\nabla_{\mu}$ is the covariant derivative, $X=-\frac{1}{2}\partial_{\mu}\phi\partial^{\mu}\phi$ is the kinetic term, $G_i(\phi, X)$ are arbitrary functions of $\phi$ and $X$, and~$_{X}$ is the derivative with respect to $X$. As~it is well known, the~Lagrangian (\ref{Horndeski}) represents the most general scalar-tensor Lagrangian that leads to second order field equations despite that it depends on second derivatives of the field $\phi$ at the level of the action as well as on non-minimally coupling terms to the Ricci scalar. As~shown in Reference~\cite{Deffayet:2011gz}, this~is just the generalisation of the so-called covariant Galileon field, the covariant version of which  loses the Galilean shift symmetry that provides its name~\cite{Deffayet:2009wt}. Hence, by~varying the action (\ref{Action}) with respect to the metric $g_{\mu\nu}$ and with respect to the scalar field $\phi$, the~corresponding field equations can be obtained and we can analyse how some particular spacetimes behave within this class of theories. 

Throughout this paper, we~are interested in studying the Nariai spacetime, which is the extremal case of the Schwarzschild-de Sitter black hole, as~is shown below. The~general Schwarzschild-de Sitter metric can be expressed in spherical coordinates as follows:
\be
ds^2=-A(r) dt^{\prime 2}+A(r)^{-1}dr^2+r^2 d\Omega_2^2\ ,
\label{SchdS}
\ee
where $d\Omega_2^2$ is the metric of a 2D sphere, and
\be
A(r)=1-\frac{2M}{r}-\frac{\Lambda}{3}r^2\ .
\label{SchdSbis}
\ee 

Here, $\Lambda>0$ and $M>0$. If~$0<M^2<\frac{1}{9\Lambda}$, the~function $A(r)$ has two positive roots $r_{BH}$ and $r_{c}$, which correspond to the black hole event horizon and to the cosmological horizon, respectively. The~global structure of this spacetime has been widely analysed in the literature~\cite{Gibbons-Hawking,Gibbons-Hawking2,Hawking:1982dh}. The~crucial point here is that whenever $M\rightarrow \frac{1}{3\sqrt{\Lambda}}$, the~size of the black hole event horizon $r_{BH}$ increases and approaches the cosmological horizon $r_c$ at $r=3M$, such~that Function (\ref{SchdSbis}) tends to:
\be
A(r)=-\frac{(r-3M)^2(r+6M)}{27M^2r}\ .
\label{SchdSbis2}
\ee

This is the extremal case of the Schwarzschild-de Sitter black hole, which is known as the Nariai spacetime~\cite{Nariai}. As~shown in (\ref{SchdSbis2}), it~leads to a degenerate horizon that corresponds to the black hole one and to the cosmological one simultaneously. The~causal structure of this particular case is well understood and the geodesics in such spacetime are well described in Reference~\cite{Podolsky:1999ts}. Note~that $A(r)\leq 0$, such~that the radial coordinate becomes timelike and the time coordinate spacelike everywhere. Our~aim here is to analyse the metric (\ref{SchdS}) for the extremal case in the framework of the Horndeski Lagrangian, and~analyse the stability of such solution. For~that purpose, let~us express the metric (\ref{SchdS}) with some more appropriate coordinates, but~firstly we express the extremal case as a limit in terms of a parameter $0<\epsilon<<1$,~\cite{Ginsparg}:
\be
9M^2\Lambda=1-3\epsilon^2\ .
\label{epsilon}
\ee

As $\epsilon\rightarrow0$, both~horizons approach each other. Then, we~can choose the following coordinates~\cite{Bousso:1997wi}:
\be
t^{\prime}=\frac{1}{\epsilon\sqrt{\Lambda}}\psi\ , \quad \quad r=\frac{1}{\sqrt{\Lambda}}\left(1-\epsilon \cos\chi-\frac{1}{6}\epsilon^2\right)\ .
\label{coordinates1}
\ee

In these new coordinates, and~expanding at first order in $\epsilon$, the~metric (\ref{SchdS}) becomes:
\be
ds^2=-\frac{1}{\Lambda}\left(1+\frac{2}{3}\epsilon\cos\chi\right)\sin^2\chi d\psi^2+\frac{1}{\Lambda}\left(1-\frac{2}{3}\epsilon\cos\chi\right)d\chi^2+\frac{1}{\Lambda}\left(1-2\epsilon\cos\chi\right)d\Omega_2^2\ .
\label{metric2}
\ee

Here the black hole horizon is given by $\chi=0$ whereas the cosmological one corresponds to $\chi=\pi$. The~spatial topology is clearly $S_1\times S_2$. By~setting $\epsilon\rightarrow 0$, the~extremal case is obtained and the metric yields (\ref{metric2}):
\be
ds^2=\frac{1}{\Lambda}\left(-\sin^2\chi d\psi^2+d\chi^2\right)+\frac{1}{\Lambda}d\Omega_2^2\ .
\label{metric3}
\ee

Finally, we~can implement another change of coordinates that simplifies the expression (\ref{metric3}), which is described by the following coordinates:
\be
x=\text{Log}\left(\tan\frac{\chi}{2}\right)\ , \quad \quad t=\frac{\psi}{4}\ .
\label{coordinates2}
\ee

The metric (\ref{metric3}) for the Nariai spacetime becomes:
\be
ds^2=\frac{1}{\Lambda\cosh^2 x}\left(-dt^2+dx^2\right)+\frac{1}{\Lambda}d\Omega_2^2\ .
\label{metric4}
\ee

The new coordinates are defined in the domain $(-\infty,\infty)$, as~can be easily shown by (\ref{coordinates2}).

\section{Reconstructing the Gravitational Action in Horndeski Gravity}
\label{solutions}

In this section, we~analyse the particular Lagrangians within Horndeski gravity that reproduces the Nariai solution. To~do so, we~use the metric as expressed in the coordinates given in (\ref{metric4}). As~shown, Nariai spacetime can be a solution for each of the Horndeski Lagrangians as far as some constraints are assumed on the $\mathcal{L}_i$ functions.

\subsection{Case with $\mathcal{L}_2$}

As a first approximation to Horndeski gravity in Nariai spacetime, we~will start studying the simplest case in which only $\mathcal{L}_2$ for $\mathcal{L}_{Hr}$ is considered,
\bea
\mathcal{L}_2=G_2(\phi,X)\label{l2}\ ,
\eea
which essentially is the usual term for K-essence theory. The~first step will be to solve, at~the background level, the~equations of motion given by the Einstein tensor plus an effective energy-tensor coming from  metric variations of the matter Lagrangian plus the Lagrangian defined in \eqref{l2}:
\bea
G_{\mu\nu}=R_{\mu\nu}-\frac{1}{2}g_{\mu\nu}R=8\pi G\left[g_{\mu\nu}G_2(\phi,X)+ \frac{\partial G_2(\phi,X)}{\partial X} \partial_\mu\phi\partial_\nu\phi+T^{(m)}_{\mu\nu}
\right]\ ,
\label{EinsteineqL2}
\eea
where $T^{(m)}_{\mu\nu}$ is the energy-momentum tensor of the matter Lagrangian, and~which, for~the case of our interest, we~are going to consider zero to focus on the vacuum, i.e., $T^{(m)}_{\mu\nu}=0$. Therefore, this~tensor equation leads to the following system of equations:
\bea
&tt-&\quad \frac{1}{8\pi G}\frac{1}{\cosh^2x}=-\frac{G_2(\phi,X)}{\Lambda\cosh^2x}+\frac{\partial G_2(\phi,X)}{\partial X}\dot\phi^2\label{tt}\ ,\\
&xt-&\quad 0=-\frac{\partial G_2(\phi,X)}{\partial X}\partial_t\phi\partial_x\phi\label{tx}\ ,\\
&xx-&\quad \frac{1}{8\pi G}\frac{-1}{\cosh^2x}=\frac{G_2(\phi,X)}{\Lambda\cosh^2x} +\frac{\partial G_2(\phi,X)}{\partial X}\phi'^2\label{xx}\ , \\
&\theta\theta-&\quad\frac{-1}{8\pi G}=\frac{G_2(\phi,X)}{\Lambda}+\frac{\partial G_2(\phi,X)}{\partial X}\partial_\theta\phi\partial_\theta\phi\label{thetatheta}\ , \\
&\Phi\Phi-& \quad\frac{-\sin^2\theta}{8\pi G}=\sin^2\theta\frac{G_2(\phi,X)}{\Lambda}+\frac{\partial G_2(\phi,X)}{\partial X}\partial_\Phi\phi\partial_\Phi\phi\ ,
\label{psipsi}
\eea
where dot means time derivatives and $'$ derivatives with respect to $x$. The~two main issues that we intend to solve are the form of $G_2(\phi,X)$ and $\phi(t,x,\theta,\Phi)$. By~combining \eqref{thetatheta} with \eqref{psipsi}, it~yields:
\bea
\partial_\Phi\phi\partial_\Phi\phi=\sin^2{\theta}\partial_\theta\phi\partial_\theta\phi\;\;\;\; \rightarrow \;\;\;\; \partial_\Phi\phi=\pm\sin{\theta}\partial_\theta\phi\ ,
\eea
the  solution of which is:
\bea
\phi=g(t,x)\left[\Phi\pm\ln\left(\cot\frac{\theta}{2}\right)\right]+f(t,x)\label{eqphi1}\ .
\eea

However, from~\eqref{tt} and \eqref{xx} it is possible to deduce that $g(t,x)$ should vanish in order to keep the same dependence parameters on the left and right hand side of the equations, and~therefore $\phi=\phi(t,x)$, which implies that $X=\Lambda\cosh^2(x)(\dot{\phi}^2-\phi'^2)/2$. This~is the formal way for showing that the scalar field has to be spherically symmetric as the metric is. In~addition, for~solving $G_2(\phi,X)$, we~can use the trace equation of \eqref{EinsteineqL2} where the scalar curvature for the Nariai metric is $R=4\Lambda$ and therefore:
\bea
-\frac{\Lambda}{4\pi G}=2 G_2(\phi,X)-\frac{\partial G_2(\phi,X)}{\partial X}X\ ,
\eea
the solution of which is:
\bea
G_2(\phi,X)=-\frac{\Lambda}{8\pi G}+f(\phi)X^2\ .
\eea

However, by Equation \eqref{tx}, the~following condition is obtained:
\bea
2Xf(\phi)\dot{\phi}\phi'=0\ .
\label{condphi}
\eea

It is straightforward to show that by combining \eqref{condphi} with $xx-$ and $tt-$ equations, $\phi'=\dot{\phi}=0$, such~that $\phi=\textit{constant}$. Hence, the~solution of the background leads to the following constraint on the~action:
\bea
G_2(\phi_0,0)=-\frac{\Lambda}{8\pi G}.
\eea

This solution mimics the one from General Relativity with a cosmological constant, but~in this case induced by a constant scalar field $\phi$. There is a special case when the coefficients for this system of equations become null and the background equation is satisfied also for non-constant and non-static scalar field solutions, which will be studied in the Appendix \ref{appendix}. Note~that despite that Birkhoff's theorem is satisfied in Brans--Dicke-like theories~\cite{Faraoni:2010rt,Schleich:2009uj,Capozziello:2011wg}, where a static metric implies a static scalar field, this~may not be the case for other scalar-tensor theories such as Galileons or general Horndeski scenarios~\cite{Babichev:2016rlq,Babichev:2016fbg}.

\subsection{Case $\mathcal{L}_3$}
\label{subsect3}
For the case $\mathcal{L}_3$, the~general gravitational action is given by
\be
S_G=\int dx^4 \sqrt{-g} \left[\frac{R}{16\pi G}-G_3(\phi, X)\Box\phi\right]\ .
\label{ActionL3}
\ee  

By varying the action (\ref{ActionL3}) with respect to the metric $g_{\mu\nu}$, the~corresponding field equations are~obtained:
\begingroup\makeatletter\def\f@size{9}\check@mathfonts
\def\maketag@@@#1{\hbox{\m@th\normalsize \normalfont#1}}%
\be
\begin{array}{rrl}
R_{\mu\nu}-\frac{1}{2}g_{\mu\nu}R&=&8\pi G\left[G_{3\phi}\left(g_{\mu\nu}\nabla_{\alpha}\phi\nabla^{\alpha}\phi-2\nabla_{\mu}\phi\nabla_{\nu}\phi\right)\right.\\
&&\left.+G_{3X}\left(-\nabla_{\mu}\phi\nabla_{\nu}\phi\Box\phi-g_{\mu\nu}\nabla_{\alpha}\phi\nabla_{\beta}\phi\nabla^{\alpha}\phi\nabla^{\beta}\phi+2\nabla^{\alpha}\phi\nabla_{(\mu}\phi\nabla_{\nu)}\nabla^{\alpha}\phi\right)\right]\ .
\label{fieldeqsL3}
\end{array}
\ee
\endgroup

Here the subscript $_{( )}$ refers to a commutator among the indexes, while $_{\phi}$ and $_{X}$ are derivatives with respect to the scalar field $\phi$ and its kinetic term $X$ respectively. The~equation for the scalar field is obtained by varying the action (\ref{ActionL3}) with respect to the scalar field:
\begingroup\makeatletter\def\f@size{8.5}\check@mathfonts
\def\maketag@@@#1{\hbox{\m@th\normalsize \normalfont#1}}%
\be
\begin{array}{lll}
&&2G_{3\phi}\Box\phi+G_{3\phi\phi}(\nabla\phi)^2+G_{3X\phi}\left[(\nabla\phi)^2\Box\phi+2\nabla_{\mu}\phi\nabla^{\mu}X\right]+G_{3X}\left[(\Box\phi)^2-\nabla_{\mu}\nabla_{\nu}\phi\nabla^{\mu}\nabla^{\nu}\phi-R_{\mu\nu}\nabla^{\mu}\nabla^{\nu}\phi\right] \\
&&+G_{3XX}\left[\nabla_{\mu}\phi\nabla^{\mu}X+(\nabla X)^2\right]=0\ ,
\label{scalarfieldEqL3}
\end{array}
\ee
\endgroup
where recall that $X$ is the kinetic term of the scalar field. As~in the previous Lagrangian, a~non-constant static scalar field, $\phi=\phi(x)$ is assumed. In~order to show that the Nariai metric, expressed in the coordinates as in (\ref{metric4}), may~be a solution for the gravitational action (\ref{ActionL3}), we~use the $tt-$ and $xx-$ equations, which can be easily obtained from the field Equations (\ref{fieldeqsL3}) and yields:
\begingroup\makeatletter\def\f@size{9}\check@mathfonts
\def\maketag@@@#1{\hbox{\m@th\normalsize \normalfont#1}}%
\be
\begin{array}{lll}
&tt-& \quad \frac{1}{\cosh^2x}=8\pi G\phi'^2\left[-G_{3\phi}+G_{3X}\Lambda^2\left(\sinh x\cosh x \phi'+\cosh^2x\phi''\right)\right] \ , \\
&xx-&-\frac{1}{\cosh^2x}=8\pi G\phi'^2\left[-G_{3\phi}+G_{3X}\Lambda^2\cosh x\sinh x \phi'\right] \ .
\label{ttxxeqsL3}
\end{array}
\ee
\endgroup

The $\theta\theta-$ and $\varphi\varphi-$ equations are just redundant, since the  $tt-$ equation is reproduced up to proportional terms. In~general, for~an arbitrary $G_3(\phi,X)$, the~system of Equation (\ref{ttxxeqsL3}) has no solution $\phi(x)$, and~consequently Nariai spacetime is not a solution for the gravitational Lagrangian (\ref{ActionL3}). Nevertheless, Equation (\ref{ttxxeqsL3}) can be used for reconstructing the appropriate $\mathcal{L}_3$ Lagrangian that reproduces the Nariai spacetime (\ref{metric4}) when assuming a particular solution $\phi(x)$. As~the corresponding partial derivatives $G_{3\phi}$ and $G_{3X}$ are at the end functions of the coordinate $x$, we~can express both of them in terms of the scalar field and its derivatives through Equation (\ref{ttxxeqsL3}), which leads to:
\begingroup\makeatletter\def\f@size{10}\check@mathfonts
\def\maketag@@@#1{\hbox{\m@th\normalsize \normalfont#1}}%
\be
\begin{array}{lll}
G_{3\phi}(x)&=&\frac{1}{8\pi G}\frac{2\tanh x\ \phi'+\phi''}{\phi^{\prime 2}\phi''\cosh^2x}\ , \\
G_{3X}(x)&=&\frac{1}{4\pi G\Lambda}\frac{1}{\phi^{\prime 2}\phi''\cosh^4x}\ .
\label{G3solutionA}
\end{array}
\ee
\endgroup

Hence, the~corresponding Lagrangian (\ref{ActionL3}) can be reconstructed as far as the expressions (\ref{G3solutionA}) are well defined for $\phi(x)$, such~that the integrability condition holds $G_{3\phi X}=G_{3X\phi}$. Nevertheless, it~is not straightforward to obtain an analytical and exact expression for the $\mathcal{L}_3$ Lagrangian, but~we can consider a couple of ways that lead to an analytical reconstruction of the action.  

Firstly, we~may specify the form of the function $G_3(\phi, X)$, and~reconstruct the corresponding action by using the system of Equation (\ref{ttxxeqsL3}) and the integrability condition on $G_3(\phi, X)$. Let~us consider the following $G_3(\phi, X)$:
\be
G_3(\phi, X)=f_1(\phi)+f_2(X)\ .
\label{G32}
\ee

The general kinetic term $X$ is given by:
\be
X=-\frac{1}{2}\Lambda\cosh^2 x\phi'^2\ .
\label{kinetic}
\ee

Then, by~the partial derivative with respect to $X$ in (\ref{G3solutionA}), we~obtain:
\be
G_{3X}=f_{2X}=\frac{\Lambda}{4\pi G}\frac{1}{X^2}\frac{\phi'^2}{\phi''}\ .
\label{f2XP}
\ee

This equation together with the assumption (\ref{G32}) basically imposes that $\frac{\phi'^2}{\phi''}=g(X)$ must be expressed as a function of the kinetic term (\ref{kinetic}). As~$g(X)$ is in principle arbitrary as far as providing a solution for the scalar field $\phi$, we~may assume $g(X)=X$ such that the scalar field becomes:
\be
\phi(x)=-\frac{2\log\left(\cosh x\right)}{\Lambda}.
\label{phiex2}
\ee

After integrating (\ref{f2XP}), the~function $f_2(X)$ turns out:
\be
f_{2}(X)=-\frac{\Lambda}{4\pi G}\frac{1}{X^2}\ .
\label{f2X}
\ee

While the partial derivative with respect to $\phi$ on $G_3$ leads to:
\be
G_{3\phi}=f_{1\phi}=\frac{\Lambda^2}{32\pi G}\frac{1+2\log(\cosh x)}{\cosh^2 x}=\frac{1}{8\pi G}\frac{\e^{\Lambda\phi}(1-\Lambda\phi)}{\phi^2}\ ,
\label{G32phi}
\ee
which after integrating, provides the corresponding dependence on the scalar field $\phi$:
\be
f_1(\phi)=-\frac{1}{8\pi G}\frac{\e^{\Lambda\phi}}{\phi}\ .
\label{f1phi}
\ee

The full gravitational action $G_3(\phi, X)$ as given in (\ref{G32}) is reconstructed. Nevertheless, we~may try to keep the form of $G_3$ arbitrary and consider a particular solution for the scalar field in order to reconstruct the action. For~illustrative purposes, we~consider the following solution:
\be
\phi(x)=\phi_0 \e^{\mu x}\ .
\label{scalarfieldEvo}
\ee

Then, by~following the Equation (\ref{G3solutionA}), the~following particular solutions are found in terms of the coordinate $x$:
\be
\begin{array}{lll}
G_{3\phi}(x)&=&\frac{1}{8\pi G}\frac{\text{sech}^2x(\mu+2\tanh x)\e^{-2\mu x}}{\mu^3\phi_0^2}\ , \\
G_{3X}(x)&=&\frac{1}{4\pi G}\frac{\text{sech}^4x\e^{-3\mu x}}{\mu^4\phi_0^3\Lambda}\ .
\label{G3solution}
\end{array}
\ee

The corresponding kinetic term $X=-\frac{1}{2}\partial_\mu\phi\partial^\mu\phi$ is given for this case by:
\be
X=-\frac{1}{2}\phi_0^2\mu^2\Lambda \cosh^2 x \e^{2\mu x}\ .
\label{kinetic1}
\ee

Hence, the~partial derivative $G_{3X}(\phi, X)$ automatically leads to:
\be
G_{3X}(\phi, X)=\frac{\Lambda}{16\pi G}\frac{\phi}{X^2}\ ,
\label{G3X1}
\ee

After integrating, it~leads to:
\be
G_{3}(\phi, X)=-\frac{\Lambda}{16\pi G}\frac{\phi}{X}+ f(\phi)\ ,
\label{G31}
\ee
where $f(\phi)$ has to be computed by integrating the partial derivative $G_{3\phi}$, which is obtained by deriving expression (\ref{G31}) and equating to the expression in (\ref{G3solution}):
\be
f_{\phi}=\frac{1}{4\pi G\phi_0^2 \mu^3}\frac{\tanh x}{\cosh^2x\e^{2\mu x}}= \frac{1}{4\pi G\phi_0^2 \mu^3}\frac{\tanh\left[\log\left(\frac{\phi}{\phi_0}\right)^{1/\mu}\right]}{\phi^2\cosh^2\left[\log\left(\frac{\phi}{\phi_0}\right)^{1/\mu}\right]}\ ,
\label{G3fp}
\ee
which after integrating, yields:
\begingroup\makeatletter\def\f@size{8}\check@mathfonts
\def\maketag@@@#1{\hbox{\m@th\normalsize \normalfont#1}}%
\be
\begin{array}{lll}
f(\phi)&=&\frac{1}{8\pi G \mu^2(\mu-2)\phi}\left\{\text{F}\left[1,1-\mu/2,2-\mu/2;-\left(\frac{\phi}{\phi_0}\right)^{2/\mu}\right]-(\mu-2)\mu\text{F}\left[1,-\mu/2,1-\mu/2;-\left(\frac{\phi}{\phi_0}\right)^{2/\mu}\right]\right. \\
 &+&\left. \text{sech}^2\left[\log\left(\frac{\phi}{\phi_0}\right)^{1/\mu}\right]+\mu\tanh\left[\log\left(\frac{\phi}{\phi_0}\right)^{1/\mu}\right]\right\}\ .
\label{G3f}
\end{array}
\ee
\endgroup

Here, $F(a,b,c;x)$ are hypergeometric functions, which can be computed analytically for some values of $\mu$. For~instance, $\mu=1$ gives:
\be
f(\phi)=-\frac{1}{4\pi G}\left[\frac{\phi^3+3\phi\phi_0^2}{(\phi^2+\phi_0^2)^2}+\frac{\arctan\left(\frac{\phi}{\phi_0}\right)}{\phi_0}\right]\ .
\label{fHyper1}
\ee

Hence, the~full reconstruction of the gravitational action (\ref{ActionL3}) is explicitly shown for these two cases. The~main conclusions can be obtained by analysing these two examples. As~shown in the field equations, and~by the expressions of $G_{3\phi}(x)$ and $G_{3 X}(x)$, a~constant scalar field $\phi(x)=\phi_0$ is not a solution for the Equation (\ref{ttxxeqsL3}), at~least whenever the Lagrangian (\ref{ActionL3}) is considered as the sole action for gravity. In~addition, the~freedom of the function $G_3(\phi, X)$ implies that different Lagrangians can reproduce the Nariai metric, but~leading to different solutions for the scalar field, as~far as its partial derivatives (\ref{G3solutionA}) are well defined, as~has been shown by these two examples.

\subsection{Case $\mathcal{L}_4$}
 
Let us now analyse the solutions when the Lagrangian $\mathcal{L}_4$ in (\ref{Horndeski}) is considered as the sole gravitational action:
\be
S_G=\int dx^4 \sqrt{-g} \left[\frac{R}{16\pi G}+G_4(\phi, X)R+G_{4X}\left((\Box\phi)^2-\nabla_\mu\nabla_\nu\phi\nabla^\mu\nabla^\nu\phi\right)\right]\ .
\label{ActionL4}
\ee  

As usual, by~varying the action (\ref{ActionL3}) with respect to the metric $g_{\mu\nu}$, the~corresponding field equations are obtained:
\begingroup\makeatletter\def\f@size{9}\check@mathfonts
\def\maketag@@@#1{\hbox{\m@th\normalsize \normalfont#1}}%
\be
\begin{array}{rrr}
\left(\frac{1}{16\pi G}+G_4\right)\left(R_{\mu\nu}-\frac{1}{2}g_{\mu\nu}R\right)-\nabla_{\mu}\nabla_{\nu}G_4+g_{\mu\nu}\Box G_4-\frac{1}{2}g_{\mu\nu}G_{4X}\left((\Box\phi)^2-\nabla_\mu\nabla_\nu\phi\nabla^\mu\nabla^\nu\phi\right)\\
+...\ \text{(second order terms)}=0\ .
 \label{fieldeqL4}
 \end{array}
 \ee
 \endgroup
 
We can proceed as in the previous Lagrangian. However, the~degree of freedom on the function $G_4(\phi, X)$ will lead to a set of infinite solutions for the scalar field, as~shown above for $\mathcal{L}_3$, which~does not provide any new insights on Nariai spacetime in Horndeski gravity, but~just some similar features as in the previous case, i.e., for~a given solution $\phi(x)$, one~can in general reconstruct the appropriate action through $G_4(\phi, X)$, while the other way around, that~is, given an arbitrary $G_4(\phi, X)$ function, the~field Equation (\ref{fieldeqL4}) does not have any solution for the scalar field in general, except for some special cases of the $G_4(\phi, X)$ function, as~also shown for $G_3(\phi, X)$ above. In~addition, note~for the general Horndeski Lagrangian, the~speed of gravitational waves is given by~\cite{Bettoni:2016mij}:
\be
c_{GW}=\frac{G_4-X(\ddot{\phi}G_{5X}+G_{5\phi})}{G_4-2XG_{4X}-X(H\dot{\phi}G_{5X}-G_{5\phi})}\ ,
\label{speedGW}
\ee
where $H$ is the Hubble parameter. Hence, by~assuming $G_4(\phi, X)=G_4(\phi)$ and $G_5=0$, analogously to~\cite{Deffayet:2010qz}, the~speed of propagation for GW's is kept as the speed of light $c_{GW}=1$, satisfying the constraints obtained from the GW170817 detection~\cite{TheLIGOScientific:2017qsa,Monitor:2017mdv}. Hence, we~explore here the case where $G_4(\phi, X)=G_4(\phi)$, such~that the field Equation (\ref{fieldeqL4}) read:
\be
\begin{array}{lll}
&tt-& \quad \left(\frac{1}{16\pi G}+G_4\right)\text{sech}^2\ x-\phi^{\prime 2}G_{4\phi\phi}-(\tanh x\phi'+\phi'')G_{4\phi}=0\ , \\
&xx-& \quad -\left(\frac{1}{16\pi G}+G_4\right)\text{sech}^2\ x-\tanh x\phi'G_{4\phi}=0\ , \\
&\theta\theta-& \quad -\left(\frac{1}{16\pi G}+G_4\right)\text{sech}^2\ x-\phi^{\prime 2}G_{4\phi\phi}+\phi''G_{4\phi}=0\ .
\label{equationsL4}
\end{array}
\ee

By combining the $xx-$ and $-\theta\theta$ equations, it~yields:
\be
\tanh x\ G_{4\phi}\phi'=0\ \rightarrow \quad \phi=\text{constant} \ . 
\label{xxthetatheta}
\ee

Hence, the~only solution leads to a constant scalar field, similarly to $G_2(\phi,X)$, unless $G_{4\phi}=0$, which together with other conditions is analysed in Appendix \ref{appendix}. For~this specific case, the~only choice of $G_4$ that satisfies the equations of motion is:
\bea
G_4(\phi)=-\frac{1}{16\pi G}\ ,
\eea

Nevertheless, for~this choice the gravitational effective coupling constant in (\ref{ActionL4}) becomes null and consequently the theory is ill defined in general. Then, for~the particular case (\ref{ActionL4}) with $G_4=G_4(\phi)$, Nariai spacetime and consequently Schwarzschild-(A)dS is not reproduced by such Lagrangian. This~is a natural consequence as Schwarzschild-(A)dS spacetime requires the presence of a cosmological constant, which cannot emerge from another term. However, such~issue can be easily sorted out by adding a scalar potential in the action, 
\be
S_G=\int dx^4 \sqrt{-g} \left[\frac{R}{16\pi G}+G_4(\phi)R-V(\phi)\right]\ .
\label{ActionL4bis}
\ee  

The equations do not differ much from the ones above, but~just up to a potential term,
\be
\begin{array}{lll}
&tt-& \quad \left(\frac{1}{16\pi G}+G_4\right)\text{sech}^2\ x-\phi^{\prime 2}G_{4\phi\phi}-(\tanh x\phi'+\phi'')G_{4\phi}-\frac{	\text{sech}^2\ x}{2\Lambda}V(\phi)=0\ , \\
&xx-& \quad -\left(\frac{1}{16\pi G}+G_4\right)\text{sech}^2\ x-\tanh x\phi'G_{4\phi}+\frac{\text{sech}^2\ x}{2\Lambda}V(\phi)=0\ , \\
&\theta\theta-& \quad -\left(\frac{1}{16\pi G}+G_4\right)\text{sech}^2\ x-\phi^{\prime 2}G_{4\phi\phi}+\phi''G_{4\phi}+\frac{\text{sech}^2\ x}{2\Lambda}V(\phi)=0\ .
\label{equationsL4bis}
\end{array}
\ee

As in the previous case, by~combining the $xx-$ and $-\theta\theta$ equations, the~constraint Equation (\ref{xxthetatheta}) is obtained, what~leads to a constant scalar field $\phi(x)=\phi_0$, and~by replacing in  Equation (\ref{equationsL4bis}), it~leads to:
\be
 -G_4(\phi_0)+\frac{V(\phi_0)}{2\Lambda}=\frac{1}{16\pi G}\ .
\label{eqpotential}
\ee

Hence, Nariai spacetime is a solution for the gravitational action (\ref{ActionL4bis}) as long as the algebraic Equation (\ref{eqpotential}) has at least a real solution. 

Therefore, it~is clear that Schwarzschild-(A)dS spacetime, and~specifically Nariai spacetime is a solution for each of the Horndeski Lagrangians whereas some constraints are imposed on the Lagrangians $\mathcal{L}_i$. It~is straightforward to show that the Nariai metric is also a solution of the full Horndeski Lagrangian as the degrees of freedom added by each $\mathcal{L}_i$ provides a way of reconstructing the corresponding gravitational action, what~will imply an infinite number of choices on the $G_i$ functions and a degenerate solution for the scalar field, as~has been shown for some of the Lagrangians above, and~which will also affect the full gravitational action due to the freedom of choosing the corresponding Lagrangians. In~the next section, we~analyse the stability of these extremal blackholes for those cases that the Nariai metric imposes real constraints on the Lagrangians.

\section{Anti-Evaporation Regime in Horndeski Gravity}
\label{antievaporation}
       

In this section, we~analyse the stability of Nariai spacetime when perturbations around the background solution are introduced. To~do so, we~focus on the first four terms of the Horndeski~Lagrangian:
\be
S_G=\int dx^4 \sqrt{-g} \left[\frac{R}{16\pi G}+G_2(\phi,X)-G_3(\phi, X)\Box\phi+G_4(\phi)R\right]\ .
\label{ActionL234}
\ee 

Note that (\ref{ActionL234}) is the most general Horndeski Lagrangian that keeps the speed of gravitational waves (\ref{speedGW}) as the speed of light. As~shown in the previous section, for~a given solution $\phi(x)$ and the Nariai metric (\ref{metric4}), one~can reconstruct the corresponding Horndeski Lagrangian that reproduces such solution. Nevertheless, here~we are assuming for simplicity while analysing the perturbations, a~constant scalar field for the background $\phi(x,t)=\phi_0$, such~that following the  results from the above section, Nariai spacetime is a solution for the gravitational action (\ref{ActionL234}) as long as the following constraint is satisfied:
\be
\frac{G_{20}}{2\Lambda}+G_{40}=-\frac{1}{16\pi G}\ .
\label{constraint234}
\ee

A useful way to define perturbations around the Nariai metric is:
\bea
ds^2=e^{2\rho(x,t)}\left(-dt^2+dx^2\right)+e^{-2\varphi(x,t)}d\Omega_2^2,
\label{metric5}
\eea
in which $\rho(x,t)$ and $\varphi(x,t)$ at the background level are: $\rho=-\ln\sqrt{\Lambda}\cosh x$ and $\varphi=\ln\sqrt{\Lambda}$. The~perturbations on the metric and the scalar field (with spherical symmetry) can be expressed as follows:
\be
\begin{array}{lll}
&&\phi \;\;\rightarrow\;\; \phi_0+\delta\phi(t,x)\\
&&\rho  \;\;\rightarrow\;\; -\ln\left[\sqrt{\Lambda}\cosh x\right]+\delta\rho\\
&&\varphi  \;\;\rightarrow\;\; \ln\sqrt{\Lambda}+\delta\varphi
\label{perturbaciones}
\end{array}
\ee

Let us show how the perturbations are transformed under a gauge transformation in order to construct gauge invariants that allow us to isolate the physical perturbations from gauge artifices. We~can consider an infinitesimal transformation of coordinates, given by
\be
x^{\prime\mu}=x^{\mu}+\delta x^{\mu}\ ,
\label{xinfi}
\ee

On any generic quantity $F$, this~implies a transformation on its perturbation:
\be
\delta F'= \delta F+ \pounds_{\delta x} F_0\ .
\label{gaugetrans}
\ee

Here the prime denotes the quantity transformed in the new coordinates, $F_0$ is the background value and $\pounds_{\delta x}$ is the Lie derivate along the vector $\delta x^{\mu}$. The~corresponding perturbations on the metric are transformed as follows:
\be
\begin{array}{rrr}
\delta\rho'=\delta\rho+\pounds_{\delta x}\rho_0\ , \\
\delta\varphi'=\delta\varphi+\pounds_{\delta x}\varphi_0=\delta\varphi\ . 
\label{gaugetrans}
\end{array}
\ee

We are interested in the perturbation $\delta\varphi$, as~the one that defines the perturbation on the radius of the horizon (see below). This~is a gauge invariant quantity, such~that we can work in an arbitrary gauge to solve the equations. Hence, introducing the perturbations (\ref{perturbaciones}) in  the field equations, up~to linear order leads to:
\be
\left(\frac{1}{16\pi G}+G_4\right)\delta G_{\mu\nu}+G_{\mu\nu}G_{4\phi}\delta\phi-G_{4\phi}\nabla_{\mu}\nabla_{\nu}\delta\phi+g_{\mu\nu}G_{4\phi}\Box\delta\phi-\frac{1}{2}\left(G_{2\phi}g_{\mu\nu}\delta\phi+G_2\delta g_{\mu\nu}\right)=0\ .
 \label{Perturfieldeq}
 \ee
 
Note that the functions $G_i$ and their derivatives are evaluated at $\phi=\phi_0$ and expanded up to first order in perturbations as follows:
\be
\begin{array}{rrr}
\left.G_2(\phi,X) \;\;\rightarrow\;\; G_2(\phi_0,0)+\frac{\partial G_2(\phi,0)}{\partial\phi}\right\vert_{\phi_0}\delta\phi \\
\left.G_4(\phi) \;\;\rightarrow\;\; G_4(\phi_0)+\frac{\partial G_4(\phi)}{\partial\phi}\right\vert_{\phi_0}\delta\phi
\label{eqXp}
\end{array}
\ee

Our next step will be the introduction of these perturbations into the field equations to study its evolution. The~$(tt)$, $(xx)$ and $(tx)$ perturbation equations are respectively:
\begingroup\makeatletter\def\f@size{8}\check@mathfonts
\def\maketag@@@#1{\hbox{\m@th\normalsize \normalfont#1}}%
\be
\begin{array}{rrr}
-2G_{20}\text{sech}^2x\delta\varphi+(G_{2\phi}+2\Lambda G_{4\phi})\text{sech}^2x\delta\phi-2G_{20}(\tanh x\delta\varphi'+\delta\varphi'')-2G_{4\phi}\Lambda(\tanh x\delta\phi'+\delta\phi'')&=&0\ , \\
-2G_{20}\text{sech}^2x\delta\varphi+(G_{2\phi}+2\Lambda G_{4\phi})\text{sech}^2x\delta\phi+2G_{20}\left(\tanh x\delta\varphi'+\delta\ddot{\varphi}\right)+2G_{4\phi}\Lambda\left(\tanh x\delta\phi'+\delta\ddot{\phi}\right)&=&0\ ,\\
G_{20}\left(\tanh x\delta\dot{\varphi}+\delta\dot{\varphi}'\right)+G_{4\phi}\Lambda\left(\tanh x\delta\dot{\phi}+\delta\dot{\phi}'\right)&=&0,
\label{perturbations}
\end{array}
\ee
\endgroup

The $(tx)-$ equation can be rewritten as follows:
\be
\begin{array}{rrr}
\frac{\partial}{\partial t}\left[G_{20}\left(\tanh x\delta\varphi+\delta\varphi'\right)+G_{4\phi}\Lambda\left(\tanh x\delta\phi+\delta\phi'\right)\right]=0\ ,\\
\rightarrow \quad  g(x,t)\tanh x+g'(x,t)=h(x)\ ,
\label{eqg}
\end{array}
\ee
where $h(x)$ is an integration function to be determined, while $g(x,t)=G_{20}\delta\varphi+G_{40}\Lambda\delta\phi$, in which integrating the Equation (\ref{eqg}) yields:
\be
g(x,t)=G_{20}\delta\varphi+G_{40}\Lambda\delta\phi=f(t)\ \text{sech}x+\text{sech}x\int \cosh x\ h(x) dx\ .
\label{expressiong}
\ee

Then, by~combining the $tt-$ and $xx-$ equations, the~functions $f(t)$ and $h(x)$ are determined,
\be
\begin{array}{rrr}
f(t)=C_1\e^t+C_2e^{-t}\ , \quad h(x)=C_3 \tanh x+C_4\ , \\
\rightarrow\quad g(x,t)=\left(C_1\e^t+C_2e^{-t}\right)\text{sech}x+C_3+C_4\tanh x\ .
\label{expressiong2}
\end{array}
\ee

Here, $C_i$'s are integration constants. Then, the~expression for the metric perturbation $\delta\varphi$ can be easily obtained:
\bea
\delta\varphi=\frac{C_1\e^t+C_2e^{-t}}{G_{20}}\text{sech}x+C_3\frac{G_{2\phi}+2\Lambda G_{4\phi}}{G_{20}(G_{2\phi}+4\Lambda G_{4\phi})}+C_4\tanh x.
\label{solPerturVarphi}
\eea

We can now calculate how the horizon changes when considering the above perturbations on the metric. The~horizon is a null hypersurface that can be defined as follows:
\be
g^{\mu\nu}\nabla_{\mu}\varphi\nabla_{\nu}\varphi=0\ ,
\label{horizon}
\ee

By introducing (\ref{perturbaciones}) and (\ref{solPerturVarphi}) in (\ref{horizon}), the~following relation is obtained:
\be
C_1^2\e^{4t}+C_2^2-\left(C_4^2+2C_1C_2\cosh2x\right)\e^{2t}+2C_1C_4\e^{3t}\sinh x+2C_2C_4\e^t\sinh x=0\ ,
\label{horizon2}
\ee
which relates the $x-$coordinate and the $t-$coordinate at the horizon:
\be
x= \log\left[\frac{C_4+\sqrt{4C_1C_2+C_4^2}}{2C_1}\e^{-t}\right]\ .
\ee

Hence, the~perturbation (\ref{solPerturVarphi}) on the metric at the horizon leads to:
\be
\delta\varphi_{h}=\frac{1}{G_{20}}\left[C_3\frac{G_{2\phi}+2\Lambda G_{4\phi}}{G_{2\phi}+4\Lambda G_{4\phi}}+\sqrt{4C_1C_2+C_4^2}\right]
\label{varphihorizon}
\ee

Therefore, the~perturbation at the horizon remains constant. By~the Nariai metric (\ref{metric5}), one~can identify the radius of the horizon when it is perturbed as:
\be
r_{h}=\frac{\e^{-\delta\varphi_h}}{\sqrt{\Lambda}}=\frac{\e^{-\frac{1}{G_{20}}\left[C_3\frac{G_{2\phi}+2\Lambda G_{4\phi}}{G_{2\phi}+4\Lambda G_{4\phi}}+\sqrt{4C_1C_2+C_4^2}\right]}}{\sqrt{\Lambda}}\ .
\label{radiohorizon}
\ee

Note that this expression is time independent, such~that no anti-evaporation effect arises when considering the restricted Horndeski Lagrangian (\ref{ActionL234}) in Nariai spacetime. The~only effect is a shift of the horizon, which may increase or decrease depending on the values of the integration constants (initial conditions) and on the functions $G_{i}$ and their derivatives evaluated at $\phi_0$  (Horndeski Lagrangian). In~addition, if~we set the integration constants to zero $C_i=0$, the~radius turns out $r_h=1/\sqrt{\Lambda}$, i.e., the~radius for the horizon in the Nariai spacetime. Moreover, by~calculating the perturbation on the scalar field $\delta\phi$ through (\ref{expressiong}), it~yields:
\be
\delta\phi(x,t)=\frac{2C_3}{G_{2\phi}+4\Lambda G_{4\phi}}\ .
\label{phiperturb}
\ee

Hence, the~scalar field perturbation does not propagate but just introduces a perturbation on the effective cosmological constant, which~explains the absence of the anti-evaporation regime and the shift of the horizon radius when considering perturbations on Nariai spacetime in the framework of Horndeski gravity.

\section{Conclusions}
\label{conclusions}

In the present paper we have analysed several aspects of Schwarzschild-de Sitter black holes, and~particularly its extremal case when both horizons, the~cosmological and the black hole ones coincide at the same hypersurface of the spacetime, the~so-called Nariai metric. Focusing on the framework of Horndeski gravity, we~have shown that the existence of such type solutions when Horndeski Lagrangians are considered can be easily achieved by the induction of an effective cosmological constant, which naturally arises when considering a constant scalar field for some of the Horndeski terms. In~addition, we~have found that not only a constant scalar field owns Nariai spacetime as a solution of the gravitational field equations but also non-constant scalar field can reproduce Schwarzschild-de Sitter extremal black holes when considering the appropriate functions on the gravitational Lagrangian. However, this~result may not satisfy the generalised Birkhoff's theorem as for Brans--Dicke-like theories~\cite{Faraoni:2010rt,Schleich:2009uj,Capozziello:2011wg}, since despite that Nariai spacetime is a static metric, the~scalar field may become non-static, as~explained in Appendix \ref{appendix}.

By considering perturbations on the background scalar field, which is assumed constant, the~induced perturbations on the metric turns out to be time dependent, which modifies the staticLK regime of the metric, inducing an exponential expansion, a~natural solution when considering an effective cosmological constant. Nevertheless, the~linear regime just induces a slight modification on the horizon radius, keeping it constant. Contrary to other frameworks where perturbations on the Nariai spacetime have been considered~\cite{Bousso:1997wi,Nojiri:1998ue,Nojiri:2013su,Sebastiani:2013fsa}, which reproduces the so-called anti-evaporation regime, where the radius of the horizon may grow with time, this~effect seems to be absent for the type of Horndeski Lagrangian analysed here. One~obviously expects to find a non-constant scalar field perturbation by going beyond the linear regime, which will consequently induce the anti-evaporation regime. In~addition, a~non-constant scalar field for the background is also expected to produce such phenomena, as~perturbations on its propagation will naturally induce effects on the horizon radius, making the Nariai metric unstable. 

\vspace{6pt} 

\acknowledgments{D.S.-C.G. is funded by the University of Valladolid (Spain). This~article is based upon work from CANTATA COST (European Cooperation in Science and Technology) action CA15117,  EU Framework Programme Horizon 2020. I.A.~is funded by Funda\c{c}\~ao para a Ci\`encia e a Tecnologia (FCT) grant number PD/BD/114435/2016 under the IDPASC PhD Program.}


\appendix

\section{}\label{appendix}

In Section \ref{solutions} we found solutions to the equations for the $\mathcal{L}_2$, $\mathcal{L}_3$ and $\mathcal{L}_4$ cases. Nevertheless, the~full set of solutions is not completely covered by the analysis above. Firstly, by~analysing the equations for the $\mathcal{L}_2$ Lagrangian given in \eqref{EinsteineqL2}, a~skillful reader could wonder about the special case in which the---in principle, non constant---coefficients of the equations become null. With~the study of this case in mind, let~us rewrite the Einstein tensor for an Einstein manifold, like~the Nariai spacetime, as: $G_{\mu\nu}=-\Lambda g_{\mu\nu}$. Therefore, Equation \eqref{EinsteineqL2} acquires the form:
\bea
0=\left[G_2(\phi,X)+\frac{\Lambda}{8\pi G}\right]g_{\mu\nu}+\frac{\partial G_2(\phi,X)}{\partial X} \partial_\mu\phi\partial_\nu\phi.
\eea

In addition, the~scalar field equation is:
\bea
0=2G_{2\phi X}(\phi,X)X-G_{2X X}(\phi,X)\nabla^\mu X\nabla_\mu\phi-G_{2X}(\phi,X)\Box\phi-G_{2\phi}(\phi,X).
\eea

So, as long as the $G_2$ function satisfies the following conditions:
\be
\begin{array}{ccc}
&G_2(\phi_0,X_0)=-\frac{\Lambda}{8\pi G}\equiv -A& \\
&G_{2X}(\phi_0,X_0)=G_{2\phi}(\phi_0,X_0)=G_{2\phi X}(\phi_0,X_0)=G_{2XX}(\phi_0,X_0)=0& \ ,
\label{constraints}
\end{array}
\ee
the field equations above hold. A~possible reconstruction of $G_2$ is given by:
\bea
G_2(\phi,X)=\sum_{n\geq 3}c_n\left[g(\phi,X)-C\right]^n-A
\eea
in which the function $g(\phi,X)$ must satisfy that of $g(\phi_0,X_0)=C$, which becomes the field equation for the scalar field. Some~examples of this can be found, but~we think that the main issue here is the possibility of solutions with $\phi_0\neq \text{constant}$. In~principle, this~fact will affect the perturbations and change our conclusions because recall that we had considered a background scalar field as a constant to solve the system of the perturbations above. For~this case, the~perturbation equations are: 
\bea
\delta\left(8\pi G\left[g_{\mu\nu}G_2(\phi,X)+ \frac{\partial G_2(\phi,X)}{\partial X} \partial_\mu\phi\partial_\nu\phi\right]\right)=\delta G_{\mu\nu}\ ,
\eea
which under the constraints \eqref{constraints} at first order yields:
\bea
\delta G_{\mu\nu}=-\Lambda \delta g_{\mu\nu}\ ,
\eea

As the perturbations on the Einstein tensor remains the same, the~following system of equations for $\delta\varphi (x,t)$ is obtained:
\be
\begin{array}{lll}
&\text{tt}-& \quad \delta\varphi''+\tanh x \delta \varphi'+\text{sech}^2x\delta\varphi=0\ , \\
&\text{xx}-& \quad \delta\ddot{\varphi}+\tanh x \delta \varphi'-\text{sech}^2x\delta\varphi=0\ , \\
&\text{tx}-& \quad \delta\dot{\varphi}'+\tanh x \delta \dot{\varphi}=0\ .
\end{array}
\ee

The general solution of this system of equations is:
\be
\delta\varphi (x,t)=\left(C_1\e^{t}+C_12e^{-t}\right)\text{sech}x + \frac{1}{2}C_3\tanh x\ ,
\ee
where $C_i$ are integration constants. Hence, the~perturbation on the horizon radius (\ref{radiohorizon}) is easily obtained:
\be
r_H=\frac{\e^{-\frac{1}{2}\sqrt{C_3^2+16C_1C_2}}}{\sqrt{\Lambda}}\ ,
\ee
which as in the case analysed throughout the paper, leads to a constant such that no instability occurs.

A similar analysis can be applied for the Lagrangians $\mathcal{L}_3$ and $\mathcal{L}_4$. For~the former, the~background equation can be expressed as:
\be
\begin{array}{lll}
-\frac{1}{2}\Lambda g_{\mu\nu}&=&8\pi G\left[G_{3\phi}\left(g_{\mu\nu}\nabla_{\alpha}\phi\nabla^{\alpha}\phi-2\nabla_{\mu}\phi\nabla_{\nu}\phi\right)\right.\\
&&\left.+G_{3X}\left(-\nabla_{\mu}\phi\nabla_{\nu}\phi\Box\phi-g_{\mu\nu}\nabla_{\alpha}\phi\nabla_{\beta}\phi\nabla^{\alpha}\phi\nabla^{\beta}\phi+2\nabla^{\alpha}\phi\nabla_{(\mu}\phi\nabla_{\nu)}\nabla^{\alpha}\phi\right)\right]\ .
\label{fieldeqsL3appendix}
\end{array}
\ee

Then, we~may impose to each term of the right hand side of this equation to be expressed as follows:
\be
\begin{array}{rrr}
G_{3\phi}\left(g_{\mu\nu}\nabla_{\alpha}\phi\nabla^{\alpha}\phi-2\nabla_{\mu}\phi\nabla_{\nu}\phi\right)= k_1 g_{\mu\nu}\ ,\\
G_{3X}\left(-\nabla_{\mu}\phi\nabla_{\nu}\phi\Box\phi-g_{\mu\nu}\nabla_{\alpha}\phi\nabla_{\beta}\phi\nabla^{\alpha}\phi\nabla^{\beta}\phi+2\nabla^{\alpha}\phi\nabla_{(\mu}\phi\nabla_{\nu)}\nabla^{\alpha}\phi\right)=k_2 g_{\mu\nu}\ ,
\end{array}
\ee
where $k_i$ are constants. Nevertheless, the~first condition leads to:
\be
\nabla_{\mu}\phi\nabla_{\nu}\phi\propto g_{\mu\nu} f(\textbf{x})\ ,
\ee
with $f(\textbf{x})$ being a function of the coordinates to be determined. By~inspecting the $tt-$ and $xx-$ equations, 
\be
\dot{\phi}^2=-\frac{1}{2}\frac{1}{\Lambda \cosh x} f(\textbf{x})\ , \quad \phi'^2=\frac{1}{2}\frac{1}{\Lambda \cosh x} f(\textbf{x})\ .
\ee

This does not guarantee a real solution for $\phi(x,t)$. Hence, we~can not find general conditions on the function $G_3$ but the system of equations has to be analysed step by step, as~done in Section \ref{subsect3}.

The case for $\mathcal{L}_4$ is similar to $\mathcal{L}_2$. The~background Equation (\ref{fieldeqL4}) hold by imposing:
\be
\begin{array}{rrl}
G_4(\phi_0,X_0)&=&-\frac{1}{16\pi G}\\
G_{4X}(\phi_0,X_0)&=&G_{4\phi}(\phi_0,X_0)=G_{4\phi X}(\phi_0,X_0)=G_{4XX}(\phi_0,X_0)=0\\
G_{4XXX}(\phi_0,X_0)&=&G_{4XX\phi}(\phi_0,X_0)=G_{4X\phi\phi}(\phi_0,X_0)=G_{4\phi\phi\phi}(\phi_0,X_0)=0 \ .
\end{array}
\ee

As above, this~can be satisfied by:
\bea
G_4(\phi,X)=\sum_{n\geq 4}c_n\left[g(\phi,X)-C\right]^n-\frac{1}{16\pi G}\ ,
\eea
where $g(\phi_0,X_0)=C$. Nevertheless, all~the coefficients of the perturbation equation become also null for any background solution $\phi_0$, such~that we have a degenerated equation that does not pose a well defined problem.


\end{document}